\documentclass[a4paper,UKenglish,cleveref, autoref, thm-restate]{lipics-v2021}

\usepackage{amsmath}
\usepackage{amssymb}
\usepackage{mathtools}
\usepackage{amsthm}
\usepackage{mathrsfs}
\usepackage{hyperref}
\nolinenumbers 

\title{From Lecture Notes to Lean: Formalizing a Textbook on Probability Theory}

\author{Shuo Deng}
{School of Data Science, The Chinese University of Hong Kong, Shenzhen, China}
{kdsdengshuo2823@gmail.com}
{https://orcid.org/0009-0002-5584-8837}
{} 

\author{Kenneth W. Shum \footnote{Corresponding author}}
{School of Science and Engineering, The Chinese University of Hong Kong, Shenzhen, China}
{wkshum@cuhk.edu.cn}
{https://orcid.org/0000-0001-6505-6177}
{Teaching Innovation Grant, Centre for Learning Enhancement and Research (CLEAR), The Chinese University of Hong Kong, Shenzhen, 2024.}

\authorrunning{Shuo Deng and Kenneth Shum}

\ccsdesc[100]{\#10003648, \#10010727} 

\keywords{Computer Formalization, Automatic Theorem Proving, Probability Theory, AI for Math, Lean.} 

\category{} 

\relatedversion{} 

\EventShortTitle{\ \ \ \ \ \ \ }

\begin{document}

\maketitle

\begin{abstract}

As large language models become increasingly capable of generating mathematical arguments, mathematics is likely to face not a scarcity of proofs but an abundance of plausible ones. In such an environment, verification, exposition, and incorporation into reusable mathematical infrastructure become central tasks. We report on an ongoing Lean formalization of \emph{Measure-Theoretic Probability: With Applications to Statistics, Finance, and Engineering}, a fourteen-chapter upper-level undergraduate textbook covering topics from Riemann--Stieltjes integration to martingales and limit theorems.

The project produces a machine-checked companion to the textbook and contributes reusable infrastructure for future formalizations involving probability theory. A Lean formalization provides computer-checked statements and proofs, makes hypotheses explicit, and allows readers to inspect the precise logical content of textbook results. A central challenge is to bridge textbook-facing statements with Mathlib's more general measure-theoretic interfaces. We reuse Mathlib results when possible and introduce reviewable interface lemmas when the textbook formulation and library abstraction differ. The project illustrates how formalized textbooks can support teaching, clarify mathematical  assumptions, and help build the formal foundations needed for reliable AI-assisted mathematics.
\end{abstract}


\section{Introduction}
\label{sec:introduction}

In the era of artificial intelligence, the teaching and communication of mathematics face both new opportunities and new challenges. Large language models can now generate mathematical explanations and proofs with increasing fluency~\cite{cao2026mechmath,feng2026towards}. However, they may also produce arguments that are logically invalid or that rely on hallucinated facts. A natural way to address this problem is to subject mathematical reasoning to independent verification by a computer proof assistant, an approach adopted for
example in~\cite{ju2604automated}. Tao has  recently argued that mathematics may be moving from an era of proof scarcity to an era of proof abundance, in which the central bottlenecks will increasingly be verification, exposition, digestion, and incorporation into the canonical theory of a field~\cite{tao2026ai}. From this perspective, formalization is not merely a way to check individual proofs; it is part of the infrastructure needed to organize mathematical knowledge in an AI-rich environment.

Among computer proof assistants, Lean has emerged as a de facto standard for the digital representation of mathematical theorems and proofs. It has been adopted, for example, in the Erd\H{o}s problems website~\cite{Bloom_Erdos_problems} and in Google's formal conjectures database~\cite{Formal_conjectures}. The usefulness of Lean depends crucially on the breadth and coherence of its mathematical library. Lean's Mathlib provides a substantial foundation for modern mathematics and already covers large parts of the undergraduate curriculum. Recent AI-assisted systems have demonstrated that, with access to this library, it is increasingly possible to formalize and prove results from undergraduate to PhD-level mathematics. For instance, contemporary large-language-model-based tools can make use of existing Mathlib theorems to produce Lean proofs of nontrivial results such as the central limit theorem. The Tau Ceti project~\cite{Tau_Ceti}, incubated by the Lean Focused Research
Organization, reflects the same broader goal: to build a collaborative and coherent library for basic mathematics. 

Some conjectures in probability theory are recently solved by large language models and verified by human mathematicians~\cite{fu2026sharp,nie2026on}. To enable automatic computer verification, there is a pressing need for systematic expansions of the formal mathematical corpus on which AI agents and human formalizers rely. For example, the works \cite{degenne2025formalization} and \cite{zhang2026statistical} formalize Brownian motion and statistical learning theory, respectively.  Further progress in AI-assisted mathematics will require formal libraries that cover more advanced theories in a reusable and pedagogically meaningful way. This is especially important for fields such as probability theory, which serves as a foundation for statistics, machine learning, mathematical finance,
information theory, and many other areas. A well-designed formalization of probability can therefore support both education and future formalization projects that depend on probabilistic reasoning.

A number of recent projects have explored the automatic formalization of mathematical textbooks and long mathematical documents. Meta's Atlas project formalizes mathematical texts at scale~\cite{rammal2026formalizing, gloeckle2026automatic}. LeanMarathon investigates long-horizon autoformalization in Lean~\cite{zhang2026leanmarathon}. We note that many such projects are primarily designed as evaluations of model capability, rather than as systematic efforts to develop the theory of a textbook into a coherent and reusable library.

There are also important examples of textbook formalization as library building. Tao initiated a Lean formalization of his textbook \emph{Analysis I}~\cite{tao_analysis_lean1}, whose careful exposition makes it a natural blueprint for formal development. That project was completed through a crowdsourced distribution of the programming effort. The second volume, \emph{Analysis II}, was subsequently translated into Lean by ReasBook~\cite{optpku2026reasbook}. Such projects illustrate how textbook formalization can serve multiple purposes: it can verify the correctness of the text, provide a machine-checkable companion for students, and contribute reusable results and interfaces to the broader formal mathematics ecosystem.

The value of such infrastructure extends beyond pure mathematics. In physics, for example, PhysLib develops reusable Lean infrastructure for formalized physics, including quantum theory and high-energy physics~\cite{physlib}. Building on this infrastructure, Meiburg {\em et al.} verified a generalized quantum Stein lemma and thereby resolved a logical gap in quantum hypothesis testing~\cite{meiburg2025formalization}. This example shows how foundational formal libraries can enable the verification of sophisticated results in downstream scientific domains.

In this paper, we document the construction of a formal companion to the measure-theoretic probability textbook~\cite{Shum2023}. Our goal is not just to test the ability of AI tools in translating isolated statements to a  computer-program format, but to develop a coherent Lean formalization aligned with the structure and pedagogy of the textbook. We first use AI agent to translate the definitions, theorems, and the proofs in Lean format. The semantic of the formalized statements are checked manually to ensure that they are aligned with the intended mathematical meaning.

The remainder of the paper is organized as follows. We first describe the scope
of the formalization and the design choices needed to align the textbook
presentation with Mathlib's measure-theoretic and probabilistic interfaces. Next, we present representative formalization cases,
including the variance of a discrete distribution and the detection of a
hidden hypothesis in a textbook theorem. We also describe the agentic workflow
used to track source passages, candidate statements, auxiliary lemmas, and
library dependencies. We
then discuss the pedagogical role of the Lean companion in an upper-level
undergraduate course.  Finally, we situate the project among related work on
textbook formalization, AI-assisted formalization, and proof assistants in
education.

\section{Scope and Interface Design}
\label{sec:mathematical-formalization}
In this paper, we document the formalization of the textbook
\emph{Measure-Theoretic Probability: With Applications to Statistics, Finance,
and Engineering}~\cite{Shum2023}, which is intended for an upper-level
undergraduate mathematics course. The scope of the project covers all fourteen
chapters of the book, including definitions, theorems, examples, and problems.
The mathematical content ranges from basic measure theory to more advanced
topics in probability, such as conditional expectation and the central limit
theorem.

We chose this textbook rather than more comprehensive standard references, such
as those by Durrett~\cite{durrett2019probability}, Klenke~\cite{klenke2008probability},
and Ross~\cite{ross2018introduction}, for both practical and pedagogical
reasons. Those texts are broader in scope and often written in a style that is
well suited to human readers but less directly aligned with formalization. By
contrast, the selected textbook belongs to the publisher's series \emph{Compact Textbooks in
Mathematics} and has a more compact structure. It is organized
around the material needed for a first course in measure-theoretic probability,
making it a more suitable target for a textbook-scale Lean companion.
The core topics contains the following items:

\begin{itemize}
\item 81 definitions,
\item 127 theorems,
\item 107 examples, and
\item 134 problems.
\end{itemize}

\noindent The formalization is currently ongoing. At this stage, the formalization of the
examples and problems has not yet been completed. The git repository can be found at
\[ \href{https://github.com/wkshum/ProbabilityTheory}{\texttt{https://github.com/wkshum/ProbabilityTheory}} 
\] 

Mathlib already provides a substantial body of measure theory and probability theory. However, its interfaces do not always follow the order, terminology, or level of abstraction used in a textbook. A central design principle of Mathlib is
to state theorems at a high level of generality, so that analogous results from different areas of mathematics can be represented uniformly and reused widely. For example, the Lebesgue integral is implemented through the Bochner integral, which is defined for functions taking values in a general Banach space; the real and complex integrals then appear as special cases.

Our project reuses Mathlib theorems directly whenever their interfaces match the textbook presentation. When they do not, we prove interface lemmas that connect the textbook-facing definitions and statements to the corresponding
library results. Such lemmas are called {\em bridges}. In this way, the project brings highly general Mathlib theorems down to a level of abstraction that is closer to the textbook and more accessible to students. This design choice is important pedagogically: measure theory and probability theory are already advanced topics for undergraduates, and the Lean statements should therefore remain as close as possible to the statements in the text. If the formal version departs too far from the textbook formulation, the Lean code becomes harder to read and less useful as a companion to the course.

Several developments in the textbook are not yet available in the current Mathlib. For example, the Vitali set is a standard example of a non-measurable
set, but this important construction does not appear to be implemented as a named object in the library. As another example, Mathlib provides infrastructure for Lebesgue--Stieltjes measures, and hence for interpreting expectations and related quantities as Lebesgue integrals with respect to such measures. However, it does not currently provide the classical Riemann--Stieltjes integral as a partition-based integral in the style used in elementary analysis.

In the textbook, Riemann and Riemann--Stieltjes integrals are used forcomputational purposes. In particular, we explain how certain Lebesgue--Stieltjes integrals can be reduced to ordinary Riemann integrals. This connection is formalized in the present project. The formalization verifies that the reduction is correct and makes the measure-theoretic Lebesgue--Stieltjes integral more accessible by relating it to the computational framework familiar
from undergraduate analysis.

\section{Representative Formalization Cases}
\label{subsec:mathematical-cases}

\subsection{Adaptation of textbook theorems to Mathlib}

Mathlib already contains a substantial collection of basic results in measure
theory and probability theory. One purpose of the present formalization is
therefore not to reprove all such results from first principles, but to explain
how theorems stated in textbook language can be connected to the corresponding
Mathlib interfaces. In this sense, the formalized textbook also functions as a
guide to the relevant parts of the Mathlib API.

A simple example is the strong law of large numbers. Mathlib provides the theorem
\[
\texttt{ProbabilityTheory.strong\_law\_ae},
\]
\noindent 
which formalizes Etemadi's version of the strong law. This version assumes that
the random variables are identically distributed and pairwise independent. It is
therefore more general than the formulation usually given in an undergraduate
probability textbook, where mutual independence is often assumed. In our
formalization, the textbook version is obtained by verifying that mutual
independence implies the pairwise independence required by Mathlib, and then
applying the existing theorem. This bridge makes explicit the relationship
between the familiar textbook statement and the more general theorem available
in the library.

A second example concerns the definition of the Lebesgue integral. In a typical
measure theory or probability textbook, the Lebesgue integral is introduced
step by step: first for simple functions, then for nonnegative measurable
functions, and finally for general real-valued measurable functions. Mathlib, by
contrast, is organized around more general integration interfaces, including
the lower Lebesgue integral for nonnegative extended-real-valued functions and
the Bochner integral for functions taking values in suitable normed spaces.
This organization is mathematically powerful, but it is not always the order or
vocabulary in which students first encounter the subject. A pedagogical goal of
the formalization is therefore to show that the textbook definitions are
consistent with, and subsumed by, the corresponding Mathlib definitions.

For instance, consider the Lebesgue integral of a nonnegative measurable
function and the monotone convergence theorem. In measure theory, a function $f$ defined on a measure space on a measure space \((\Omega,\mathscr{F},\mu)\)  is said to be  {\em simple} if it is measurable and the range of $f$ is finite. In general, we can write $f$ as finite sum of indicator function 
$$
f = \sum_{i=1}^n a_i \mathbf{1}_{A_i},
$$
where $\mathbf{1}_A$ is the indicator function of $A$, $n$ is a positive integer, and $A_i$'s are mutually disjoint measurable sets. The Lebesgue integral of $f$ is then  defined as a finite sum
$$
\int f \, d\mu \triangleq \sum_{i=1}^n a_i \mu(A_i)
$$
Using the integral of simple functions, the integral of a nonnegative measurable
function \(X\) is defined as
\[
  \int X\,d\mu
  \triangleq
  \sup\left\{
    \int f\,d\mu :
    f \text{ is simple and } 0 \leq f \leq X
  \right\}.
\]
Here \(f\) ranges over all nonnegative simple functions bounded above by \(X\).
The monotone convergence theorem is then proved soon after this definition.

In the Lean formalization, we first prove that this textbook definition agrees
with Mathlib's lower Lebesgue integral, represented by \texttt{lintegral}.
After this identification has been established, the monotone convergence
theorem can be obtained by invoking the existing Mathlib theorem
\[
\texttt{MeasureTheory.lintegral\_iSup}.
\]
\noindent 
This example illustrates the role of interface lemmas in the project. Rather than duplicating Mathlib's development of integration theory, we provide bridges from the textbook presentation to the library API. These bridges help
students and future users understand how the more abstract theorem in Mathlib corresponds to the formulation found in a standard probability text.

\subsection{Variance of a discrete probability distribution}

One example in the textbook defines the variance of a discrete random variable
with probability mass function \(p_i\) and finite mean \(m\) by the direct
formula
\[
  \operatorname{Var}(X)
    = E[(X-m)^2]
    = \sum_i (i-m)^2 p_i,
\]
\noindent 
without first constructing the distribution of the transformed random variable
\(Y=(X-m)^2\). An initial Lean formulation followed this route and compiled, but
it exposed the mean and variance through Mathlib's integral interface without
carrying the corresponding integrability hypotheses.

This issue reflects an important difference between informal mathematics and
formalization in Lean. In Lean, functions are total: a function must return a
value for every element of its domain. Consequently, an expression representing
the expectation of a non-integrable function may still have a fallback value,
rather than being undefined as a real-valued mathematical expectation. The
resulting declaration was therefore technically valid, but it imposed weaker
domain discipline than the textbook's finite-mean assumption requires.

The repaired interface makes the witness for the mean explicit through the Mathlib's API
\texttt{IsDiscretePmfMean}, and it requires integrability of the centered square
before exposing the variance as a real number. The proof then applies
\texttt{PMF.integral\_eq\_tsum} and rewrites scalar multiplication in order to
recover the textbook sum. This example illustrates a recurring translation
principle in the project: a total function provided by the library should be used
as a mathematical quantity only after the intended domain conditions have been
made explicit.

\subsection{Revealing an error in the textbook}

The formalization also helped identify and correct a mistake in the textbook
concerning the Riemann--Stieltjes integral. Theorem~1.2 asserts the additivity
formula
\[
\int_a^b f \, d\alpha =
\int_a^c f \, d\alpha +
\int_c^b f \, d\alpha ,
\]

\noindent 
where \(a<c<b\), under the assumption that \(f\) is integrable on the two
subintervals \([a,c]\) and \([c,b]\). The theorem then incorrectly concludes
that \(f\) is integrable on the whole interval \([a,b]\).

During formalization, the AI agent flagged this statement and produced a
counterexample. In the context of the textbook, the integrability of \(f\) on
\([a,b]\) is in fact guaranteed by a standing hypothesis on the integrand \(f\)
and the integrator \(\alpha\). However, that hypothesis is not included in the
statement of Theorem~1.2 itself, and the conclusion is therefore not valid as
stated.

This example illustrates one of the benefits of formalization: it forces hidden
assumptions to be made explicit. Even when an informal argument is correct in
the surrounding mathematical context, the formal statement must contain all
hypotheses needed for the theorem to hold.

\section{Agentic AI-Assisted Formalization}
\label{sec:agentic-ai}

The formalization was carried out with the assistance of an agentic workflow, with inspiration from the agentic framework APOLLO (Automated PrOof repair viaLLM and Lean cOllaboration) proposed in~\cite{NEURIPS2025_3b77109a}.
At a high level, the project distinguishes three kinds of artifacts. The
\emph{source artifact} consists of the textbook excerpts assigned to individual
formalization tasks, together with any later corrections or errata. The
\emph{mathematical artifact} consists of the Lean files, their imports, and the
pinned Lean/Mathlib environment in which they are checked. The
\emph{process artifact} records the history of the formalization: candidate
statements, auxiliary lemmas, build results, review requests, review verdicts,
and later repairs. This separation is important because these artifacts answer
different questions. Lean checks whether a formal proof is valid in the given
environment; it does not by itself check whether the formal statement is the
intended translation of the textbook.

The workflow begins by converting selected parts of the textbook into source-bound tasks. Each task retains the relevant source passage, its place in the mathematical order of the text, and its declared dependencies. The agent then proposes Lean statements and proofs, using existing Mathlib results whenever possible and introducing project-local bridge lemmas when the textbook interface does not match the library interface. The complete task include a
public theorem file and supporting helper files owned by the task. This is important in practice: a small public statement may depend on substantial task-specific  infrastructure, and reviewing only the final theorem can miss the mathematical content of the formalization.

After a candidate is produced, the Lean build checks that the relevant files compile and that the project satisfies its admission and axiom checks. This is the formal proof-checking stage. A successful build establishes that Lean
accepts the candidate, but it does not establish source fidelity. The second stage is therefore a read-only review of the candidate against the textbook source. The reviewer is asked to identify the hypotheses and conclusions of the
source statement, locate their counterparts in Lean, and check whether the formal theorem proves the intended claim rather than a weakened, strengthened, or unrelated version of it. The review also inspects task-owned helper lemmas
and relevant library interfaces when they affect the mathematical claim.

This review is designed to detect several common failure modes in AI-assisted formalization. A candidate may prove only a special case of the textbook theorem, add an extra hypothesis that hides the difficult part of the argument, replace an identity by an existence statement, or wrap a stronger library theorem in a way that bypasses the mathematical route intended by the text. A helper lemma may also move the main obligation into an assumption rather
than proving it. 
For this reason, source-fidelity review is treated as a separate layer of evidence.

The final status of a task is updated only after both stages have been
considered. A favorable review is necessary for clean completion, but it is not
automatically sufficient. The application step checks that the reviewed source,
candidate, helper files, and relevant dependencies are still current. If the
source passage has been corrected, if the Lean statement has changed, or if a
relevant helper or library interface has been modified, the previous review is
treated as stale and must be refreshed. Failed attempts are retained in the
project history rather than overwritten. Thus the record can show not only the
current accepted formalization, but also why earlier candidates failed and what
changed in the repair.

When a review fails, the workflow routes the failure according to its nature. A
local proof gap may return to ordinary proof repair. A mismatch between the
source statement and the Lean statement may require redesigning the public
interface. A suspected error in the textbook is escalated to the author and, if
confirmed, recorded as a source correction. A result that the textbook cites
from outside the course may be marked as an explicit scope boundary rather than
as a completed internal proof. In each case, a later accepted version must pass
again through build, review, and application; earlier evidence is not reused for
a changed task.

This architecture reflects the main methodological position of the project. Compilation, absence of \texttt{sorry}, and a controlled axiom footprint are essential, but they are not enough to establish faithful textbook formalization. 
Human judgment remains necessary in deciding whether a theorem statement has the right mathematical content, whether an abstraction preserves the intent of the textbook, and whether a suspected source error should be corrected. The value of the workflow is therefore procedural rather than absolute: it makes these
judgments explicit, records the evidence on which they depend, and keeps the history of failures and repairs inspectable.


\section{Related Work}
\label{sec:related-work}

A closely related methodological comparison is Coelho's Lean development of
mathematical finance~\cite{coelho2026finance}. That project builds a reusable
domain-specific library containing more than 300 sorry-free theorems across
eleven areas of mathematical finance. It also tracks the status of 330 results,
classifying each as a full formalization of the intended claim, a thin wrapper
around an existing library theorem, a reduced core statement with additional
assumptions, or a placeholder. In addition, the project uses a build-enforced
gate to control the axioms on which the formalized theorems depend. These design
choices are close in spirit to ours: both projects emphasize  the development of reviewable, reusable infrastructure for a specific mathematical domain.

Several recent projects have focused on large-scale or automated textbook
formalization. Gloeckle {\em et al.} report a system that produced approximately
130{,}000 lines of Lean code and 5{,}900 declarations from a graduate
textbook~\cite{gloeckle2026automatic}. M2F describes a two-stage pipeline
applied to 479 pages of real and convex analysis~\cite{wang2026m2f}. Similarly,
AutoformBot/ATLAS reports more than 45{,}000 declarations drawn from 26
textbooks~\cite{rammal2026formalizing}. These systems incorporate mechanisms
such as provenance tracking, human review, and persistent project state. Their
main emphasis, however, is on scaling autoformalization pipelines and evaluating
the capabilities of AI systems. By contrast, the present project is organized
around a textbook-facing development of probability theory, with particular
attention to the interfaces between the exposition in the book and the existing
abstractions in Mathlib.

Recent experience reports also identify an important boundary between automated
formalization and mathematical judgment. Ilin documents issues such as
hypothesis creep, failures of definition alignment, prompt design, commit
history, and human review in a semi-autonomous research formalization
project~\cite{ilin2026semi}. Miller makes several formal quality criteria
machine-checkable, including successful compilation, absence of
\texttt{sorry}, axiom footprint, and the export of a reusable layer, while
leaving the judgment that the formal statement matches the intended theorem to
the mathematician~\cite{miller2026vlasov}. Similarly, LeanMarathon and Automatic
Textbook Formalization study show that structural checks and local proof correctness can coexist with a wrong abstraction, or a specification that is mathematically unrelated to
the intended statement~\cite{gloeckle2026automatic, zhang2026leanmarathon}.

There is also a growing literature on the use of proof assistants in education.
A recent survey documents theorem provers and pedagogical adaptations of general
proof assistants for teaching logic, mathematics, and computer
science~\cite{minh2025proof}. Comparative work shows that proof assistants
differ substantially in their languages, interaction models, automation, and
proof presentation. As a result, educational claims about one system or course
design cannot be transferred uncritically to another~\cite{bartzia2026proof}.
For Lean specifically, an exploratory undergraduate study reports that students
can follow multiple viable paths in proof construction, despite the demands of
formal syntax~\cite{hanna2024using}. Work on Waterproof studies a system built
around controlled natural language and reports suggestive evidence of transfer
to paper proof construction, while also noting that self-selection prevents a
causal interpretation of the observed performance differences~\cite{otte2026educational}.

\section{Discussion}
\label{sec:discussion}

The formalization accompanies the textbook in several ways. Readers can inspect the exact hypotheses of a result, experiment with variants, and receive kernel-checked feedback. Some theorems are proved informally in lectures, with certain details omitted. These details are often conveyed by drawing a diagram, working through a representative example, or appealing to symmetry. When a student wants to see how the full proof is carried out, they can refer to the corresponding Lean proof. It may be surprising to observe that an idea that is simple and transparent to a human reader can require thousands of lines of code to justify in a proof assistant.

The project also clarifies the role of Mathlib. Reuse is most reliable when the textbook and the library expose the same mathematical interface. When they do not, a reviewable bridge is preferable either to re-implementing an existing library theorem or to silently altering the textbook statement. Such a bridge is itself part of the mathematical explanation: it records exactly what must be established before a library result can discharge a textbook obligation. In this sense, the formalized textbook can also be regarded as a guide to using the APIs provided by Lean's mathematical library. It shows how the material in the textbook is compatible with the theorems already available in Mathlib, while
making explicit the translations and auxiliary results needed to connect the two.  In this way, this project contributes both to the formalization of probability theory and to the development of proof-assistant resources for advanced undergraduate teaching.

\section{Acknowledgements}
The author gratefully acknowledges the assistance of several AI tools during
the preparation of this formalization project and manuscript, including Gemini,
GPT-5.5, and Harmonic Aristotle. These tools were used to support tasks such as
drafting, proof search, and the refinement of formalization
strategies. All mathematical statements, Lean developments, and expository
claims reported in this paper were reviewed and approved by the author.

\section{Conclusion}
\label{sec:conclusion}

We have presented an ongoing Lean formalization of a measure-theoretic
probability textbook, with the goal of producing both a machine-checked
companion for teaching and reusable infrastructure for future developments in
probability theory. The project covers a broad range of material, from
Riemann--Stieltjes integration to conditional expectation and the central limit
theorem. A central lesson of the project is that formalizing a textbook is not a
mechanical transcription of prose into Lean. It requires reconstructing
mathematical interfaces, and deciding how textbook definitions, hypotheses,
examples, and proof patterns should be represented in a formal library whose
abstractions may differ substantially from the exposition in the book.

This distinction also clarifies the meaning of machine-checked correctness.
Lean verifies a formal statement relative to its definitions, hypotheses,
imported library, and axioms. It does not, by itself, certify that the formal
statement is the intended translation of the corresponding textbook statement.
Successful compilation is therefore necessary, but not sufficient, evidence of
faithful formalization.

The agentic workflow used in the project makes this additional layer of
judgment inspectable. Each formalization task is tied to its source passage, the
candidate Lean statement, the auxiliary lemmas introduced for the project, and
the library results on which it depends. When a candidate fails, the failure is
not erased from the record; it remains part of the history explaining why a
repair, a change in formulation, or a correction to the source text was needed.
This record makes it possible to distinguish a Lean artifact that is merely
valid from one that faithfully represents the intended textbook mathematics.

This paper reports on a single formalization project, and its conclusions should
therefore be interpreted accordingly. We do not claim to measure the reliability
of AI reviewers or to establish the superiority of an agentic workflow over
traditional human formalization. Rather, our contribution is to document, in a
concrete probability-theory development, where a valid Lean artifact and a
faithful textbook translation can diverge, and how such divergences can be
detected and repaired.

From a pedagogical perspective, the project shows that a formalized textbook can
serve as more than a repository of checked proofs. It can function as an
interactive companion to a difficult course, helping students inspect omitted
details and helping instructors detect hidden hypotheses in lecture notes and
textbook statements. This role is especially important in an AI-rich teaching
environment, where students can easily generate plausible proofs but still need
tools for verifying their correctness. As AI systems make the
generation of mathematical arguments increasingly easy, verification and
understanding will become more central. Computer formalization is therefore
likely to play an increasingly important role in both mathematical research and
mathematics education.


\begin{thebibliography}{10}
	
	\bibitem{bartzia2026proof}
	Evmorfia-Iro Bartzia, Emmanuel Beffara, Antoine Meyer, and Julien Narboux.
	\newblock Proof assistants for undergraduate mathematics education: Elements of
	an a priori analysis.
	\newblock {\em Int. J. of Mathematical Education in Science and Technology},
	pages 1--44, 2026.
	\newblock \href {https://doi.org/10.1080/0020739X.2026.2632264}
	{\path{doi:10.1080/0020739X.2026.2632264}}.
	
	\bibitem{Bloom_Erdos_problems}
	Thomas Bloom.
	\newblock Erdos problems website.
	\newblock \href{https://www.erdosproblems.com/}{link to webpage}, 2023.
	
	\bibitem{cao2026mechmath}
	Yichuan Cao, Ruichen Qiu, Junqi Liu, Jiaqi Wang, Dakai Guo, Ruyong Feng, Lihong
	Zhi, and Xiao-Shan Gao.
	\newblock Mech{M}ath agent team: {LLM} driven agents for mathematical research.
	\newblock \href{https://arxiv.org/abs/2607.04394}{arXiv:2607.04394 [cs.AI]},
	July 2026.
	
	\bibitem{coelho2026finance}
	Raphael Coelho.
	\newblock A formally verified library of mathematical finance in {L}ean 4.
	\newblock \href{https://arxiv.org/abs/2606.01356v3}{arXiv:2606.01356v3
		[cs.LO]}, 2026.
	
	\bibitem{degenne2025formalization}
	Remy Degenne, David Ledvinka, Etienne Marion, and Peter Pfaffelhuber.
	\newblock Formalization of {B}rownian motion in {L}ean.
	\newblock \href{https://arxiv.org/abs/2511.20118}{ arXiv:2511.20118 [math.PR]},
	December 2025.
	
	\bibitem{durrett2019probability}
	Rick Durrett.
	\newblock {\em Probability: theory and examples}, volume~49.
	\newblock Cambridge university press, 2019.
	
	\bibitem{feng2026towards}
	Tony Feng, Trieu~H Trinh, Garrett Bingham, Dawsen Hwang, Yuri Chervonyi,
	Junehyuk Jung, Joonkyung Lee, Carlo Pagano, Sang-hyun Kim, Federico
	Pasqualotto, et~al.
	\newblock Towards autonomous mathematics research.
	\newblock \href{https://arxiv.org/abs/2602.10177}{arXiv:2602.10177 [cs.LG]},
	March 2026.
	
	\bibitem{Tau_Ceti}
	Lean {FRO}.
	\newblock Tau {C}eti project.
	\newblock \href{https://github.com/TauCetiProject/TauCeti}{link to github
		repo}, 2026.
	
	\bibitem{fu2026sharp}
	Weibo Fu, Yanjun Han, Guanyang Wang, Jun Yan, Peng Zhang, and Zhengqing Zhou.
	\newblock Sharp small-deviation inequalities for sums of independent
	nonnegative random variables.
	\newblock \href{https://arxiv.org/abs/2607.23980}{ arXiv:2607.23980 [math.PR]},
	July 2026.
	
	\bibitem{gloeckle2026automatic}
	Fabian Gloeckle, Ahmad Rammal, Charles Arnal, Remi Munos, Vivien Cabannes,
	Gabriel Synnaeve, and Amaury Hayat.
	\newblock Automatic textbook formalization.
	\newblock \href{https://arxiv.org/abs/2604.03071}{arXiv:2604.03071 [cs.AI]},
	2026.
	
	\bibitem{Formal_conjectures}
	{Google Deepmind}.
	\newblock Formal conjectures.
	\newblock \href{https://google-deepmind.github.io/formal-conjectures/}{link to
		webpage}.
	
	\bibitem{hanna2024using}
	Gila Hanna, Brendan Larvor, and Xiaoheng~Kitty Yan.
	\newblock Using the proof assistant {L}ean in undergraduate mathematics
	classrooms.
	\newblock {\em ZDM--Mathematics Education}, 56:1517--1529, 2024.
	\newblock \href {https://doi.org/10.1007/s11858-024-01577-9}
	{\path{doi:10.1007/s11858-024-01577-9}}.
	
	\bibitem{ilin2026semi}
	Vasily Ilin.
	\newblock Semi-autonomous formalization of the {V}lasov-{M}axwell-{L}andau
	equilibrium.
	\newblock \href{https://arxiv.org/abs/2603.15929}{arXiv:2603.15929 [cs.AI]},
	March 2026.
	
	\bibitem{ju2604automated}
	Haocheng Ju, Guoxiong Gao, Jiedong Jiang, Bin Wu, Zeming Sun, Leheng Chen,
	Yutong Wang, Yuefeng Wang, Zichen Wang, Wanyi He, et~al.
	\newblock Automated conjecture resolution with formal verification.
	\newblock \href{https://arxiv.org/abs/2604.03789}{arXiv:2604.03789v2 [cs.LG]},
	May 2026.
	
	\bibitem{klenke2008probability}
	Achim Klenke.
	\newblock {\em Probability theory: a comprehensive course}.
	\newblock Springer, 2008.
	
	\bibitem{meiburg2025formalization}
	Alex Meiburg, Leonardo~A Lessa, and Rodolfo~R Soldati.
	\newblock A formalization of the generalized quantum {S}tein's lemma in {L}ean.
	\newblock \href{https://arxiv.org/abs/2510.08672}{arXiv:2510.08672 [quant-ph]},
	October 2025.
	
	\bibitem{miller2026vlasov}
	Joseph~K. Miller.
	\newblock A formalization of the mean-field derivation of the {V}lasov
	equation.
	\newblock \href{https://arxiv.org/abs/2607.08986v2}{arXiv:2607.08986v2
		[cs.AI]}, 2026.
	
	\bibitem{minh2025proof}
	Fr{\'e}d{\'e}ric~Tran Minh, Laure Gonnord, and Julien Narboux.
	\newblock Proof assistants for teaching: a survey.
	\newblock \href{https://arxiv.org/abs/2505.13472}{arXiv:2505.13472 [cs.LO]},
	2025.
	
	\bibitem{nie2026on}
	Zipei Nie and Jiaye Wei.
	\newblock On {F}eige's conjecture.
	\newblock \href{https://arxiv.org/abs/2607.24528}{ arXiv:2607.24528 [math.PR]},
	July 2026.
	
	\bibitem{NEURIPS2025_3b77109a}
	Azim Ospanov, Farzan Farnia, and Roozbeh Mohit.
	\newblock {APOLLO}: Automated {LLM} and {L}ean collaboration for advanced
	formal reasoning.
	\newblock In {\em Advances in Neural Information Processing Systems (NuerIPS)},
	volume~38, pages 41599--41633, 2025.
	
	\bibitem{otte2026educational}
	Pim Otte, Rogier Bos, Johan Commelin, and Jim Portegies.
	\newblock The educational proof assistant {W}aterproof in an introductory proof
	course: Proof construction and learning processes.
	\newblock \href{https://arxiv.org/abs/2606.26809}{arXiv:2606.26809 [math.HO]},
	2026.
	
	\bibitem{rammal2026formalizing}
	Ahmad Rammal, Niket Patel, Fabian Gloeckle, Amaury Hayat, Julia Kempe, Remi
	Munos, Charles Arnal, and Vivien Cabannes.
	\newblock Formalizing mathematics at scale.
	\newblock \href{https://arxiv.org/abs/2605.29955}{arXiv:2605.29955 [cs.AI]},
	May 2026.
	
	\bibitem{optpku2026reasbook}
	{ReasBook contributors}.
	\newblock {ReasBook}: Lean 4 formalizations of mathematics from textbooks and
	research papers.
	\newblock \url{https://github.com/optpku/ReasBook/}, 2026.
	
	\bibitem{ross2018introduction}
	Sheldon~M Ross.
	\newblock {\em Introduction to probability models}.
	\newblock Academic press, 10th edition, 2018.
	
	\bibitem{Shum2023}
	Kenneth~W. Shum.
	\newblock {\em Measure-theoretic Probability -- with Applications to
		Statistics, Finance and Engineering}.
	\newblock Birkh{\"a}user, 2023.
	\newblock \href {https://doi.org/10.1007/978-3-031-49830-5}
	{\path{doi:10.1007/978-3-031-49830-5}}.
	
	\bibitem{tao_analysis_lean1}
	Terence Tao.
	\newblock Lean companion to {A}nalysis {I}.
	\newblock \url{https://github.com/teorth/analysis}, 2025.
	\newblock GitHub repository.
	
	\bibitem{tao2026ai}
	Terence Tao.
	\newblock Mathematics in the age of {AI}.
	\newblock Public lecture, International Congress of Mathematicians (ICM), July
	2026.
	
	\bibitem{physlib}
	{The Phys{L}ib community}.
	\newblock Phys{L}ib: The {L}ean physics library.
	\newblock \url{https://github.com/leanprover-community/physlib}, 2024.
	\newblock GitHub repository, accessed July 23, 2026.
	
	\bibitem{wang2026m2f}
	Zichen Wang, Wanli Ma, Zhenyu Ming, Gong Zhang, Kun Yuan, and Zaiwen Wen.
	\newblock {M2F}: Automated formalization of mathematical literature at scale.
	\newblock \href{https://arxiv.org/abs/2602.17016}{arXiv:2602.17016 [cs.AI]},
	2026.
	
	\bibitem{zhang2026statistical}
	Yuanhe Zhang, Jason~D. Lee, and Fanghui Liu.
	\newblock Statistical learning theory in {L}ean 4: Empirical processes from
	scratch.
	\newblock \href{https://arxiv.org/abs/2602.02285v1}{ arXiv:2602.02285 [cs.LG]},
	February 2026.
	
	\bibitem{zhang2026leanmarathon}
	Yuanhe Zhang, Yuekai Sun, Taiji Suzuki, Jason~D Lee, and Fanghui Liu.
	\newblock Lean{M}arathon: Toward reliable {AI} co-mathematicians through
	long-horizon {L}ean autoformalization.
	\newblock \href{https://arxiv.org/abs/2606.05400}{arXiv:2606.05400 [cs.AI]},
	2026.
	
\end{thebibliography}

\end{document}